\documentclass[
    ,final            
  ]
  {aipproc}
\layoutstyle{6x9}
\usepackage{epsfig}
\newcommand{\ben}{\begin{eqnarray}}
\newcommand{\een}{\end{eqnarray}}
\newcommand{\bef}{\begin{figure}[htb]\centering}
\newcommand{\eef}{\end{figure}}
\newcommand{\nnu}{\nonumber\\}

\begin{document}

\title{Trigluon correlations and single transverse spin asymmetry in open charm production}


\author{Zhong-Bo Kang}{
  address={Department of Physics and Astronomy,
                 Iowa State University,
                 Ames, IA 50011, USA}}

\author{Jian-Wei Qiu}{
  address={Department of Physics and Astronomy,
                 Iowa State University,
                 Ames, IA 50011, USA}}

\begin{abstract}
We study the single transverse-spin asymmetry for open charm production in the semi-inclusive lepton-hadron deep inelastic scattering (SIDIS) and $pp$ collision. Within collinear factorization approach, we find that the asymmetry is sensitive to the twist-3 trigluon correlation functions in the proton. With a simple model for the trigluon correlation functions, we estimate the asymmetry in SIDIS for both COMPASS and eRHIC kinematics, as well as in $pp$ collision at RHIC energy.  
We discuss the possibilities of extracting the trigluon correlation functions in these experiments.
\end{abstract}

\maketitle


\section{Introduction}

Large single transverse-spin asymmetries (SSAs)
have been consistently observed in various experiments 
at different collision energies. 
From the parity and time-reversal invariance of 
the strong interaction dynamics, the measured large asymmetries 
in high energy collisions should be directly connected to 
the transverse motion of partons inside a polarized hadron. 
Understanding the QCD dynamics behind the measured 
asymmetries should have the profound impact on 
our knowledge of strong interaction and hadron structure. 

Within the QCD collinear factorization formalism for the cross 
sections with a large momentum transfer, the observed phenomenon 
of SSAs are attributed to three-parton correlations
inside a polarized hadron. For the leading order gluonic
pole contribution to the SSAs \cite{Efremov,qiu},
the relevant partonic correlations are given by the twist-3
quark-gluon correlation function 
\ben
T_{q, F}(x, x)=\int\frac{dy_1^- dy_2^-}{4\pi}e^{ixP^+y_1^-}
\langle P,s_T|\bar{\psi}_q(0)\gamma^+\left[ \epsilon^{s_T\sigma n\bar{n}}F_\sigma^{~ +}(y_2^-)\right] \psi_q(y_1^-)|P,s_T\rangle,
\label{Tq}
\een
and by a twist-3 trigluon correlation function 
\cite{Ji:1992eu,Kang:2008qh,Kang:2008ih},
\ben
T_G(x, x)=\int\frac{dy_1^- dy_2^-}{2\pi}e^{ixP^+y_1^-}\frac{1}{xP^+}\langle P,s_T|F^+_{~~\alpha}(0)\left[ \epsilon^{s_T\sigma n\bar{n}}F_\sigma^{~ +}(y_2^-)\right] F^{\alpha+}(y_1^-)|P,s_T\rangle .
\label{Tg}
\een
There is
a quark-gluon correlation function, $T_{q,F}$, for each quark (anti-quark) flavor $q$
($\bar{q}$), and there are two independent trigluon correlation functions,
$T_G^{(f)}(x, x)$ and $T_G^{(d)}(x, x)$,
because of the fact that the color of the three gluon field strengths in Eq.~(\ref{Tg}) can be neutralized by contracting with either the antisymmetric $if^{ABC}$ or the symmetric $d^{ABC}$ tensors with color indices, $A$, $B$, and $C$
\cite{Ji:1992eu,Kang:2008qh,Kang:2008ih}.
For the high order gluonic pole contribution as well as the
fermionic pole contribution \cite{Efremov,qiu}, 
another set of quark-gluon and trigluon
correlation functions is also relevant \cite{Kang:2008ey}.

In this talk, we present the calculation of the leading gluonic
pole contribution to the SSA of open charm production in both the 
SIDIS and $pp$ collisions.  We explore the role of the trigluon
correlation functions in generating the SSA of $D$ (or $\bar{D}$) 
mesons.

\section{Single Transverse-Spin Asymmetry in SIDIS and $pp$ collision}

We first study the SSAs in SIDIS, 
$e(\ell)+p(P, s_\perp)\to e(\ell')+h(P_h)+X$, 
with $s_\perp$ the transverse spin vector of the initial hadron
and $h$ the observed $D$ 
or $\bar{D}$ meson .
We work in the approximation of one-photon exchange, and
 define the virtual photon momentum $q=\ell-\ell'$ and 
 its invariant mass $Q^2=-q^2$. We adopt the usual SIDIS variables:
\ben
S_{ep}=(P+\ell)^2, \qquad 
x_B=\frac{Q^2}{2P\cdot q},\qquad
y=\frac{P\cdot q}{P\cdot \ell}=\frac{Q^2}{x_B S_{ep}},\qquad 
z_h=\frac{P\cdot P_h}{P\cdot q}.
\een
The single transverse-spin asymmetry is defined as
\ben
A_N=\frac{\sigma(s_\perp)-\sigma(-s_\perp)}{\sigma(s_\perp)+\sigma(-s_\perp)}
=\frac{d\Delta\sigma(s_\perp)}{dx_B dy dz_h dP_{h\perp}^2 d\phi}\left/
\frac{d\sigma}{dx_B dy dz_h dP_{h\perp}^2 d\phi}\right.,
\label{an_sidis}
\een
where $\phi$ is the azimuthal angle between the hadron plane 
defined by initial- and final-hadron and the lepton plane 
defined by the initial- and final-lepton. 

Within the collinear factorization formalism, 
the leading order contribution for open charm
production comes from the photon-gluon fusion subprocess at the partonic level.
The single transverse-spin-dependent cross section for $D$ meson production 
can be written as
\ben
\frac{d\Delta\sigma(s_\perp)}{dx_B dy dz_h dP_{h\perp}^2 d\phi}
&=&
\sigma_0\int_{x_{min}}^1 dx\int \frac{dz}{z}D(z)
\delta\left(\frac{P_{h\perp}^2}{z_h^2}
           -\frac{(1-\hat{x})(1-\hat{z})}{\hat{x}\hat{z}}Q^2
           +\hat{z}^2 m_c^2\right)\nnu
&\times&\left(\frac{1}{4}\right) 
\left[\epsilon^{P_h s_\perp n \bar{n}}
\left(\frac{\sqrt{4\pi\alpha_s}}{z\hat{t}}\right)
\left(1+\frac{\hat{t}}{\hat{u}}\right)\right]
\nnu
&\times&
\sum_{j=f,d}\sum_{i=1}^{4} {\cal A}_i 
\left[-x\frac{d}{d x}\left(\frac{T_G^{(j)}(x,x)}{x}\right)\hat{W}_i+\left(\frac{T_G^{(j)}(x,x)}{x}\right)\hat{N}_i\right],
\label{dmeson}
\een
where $\sigma_0=e_c^2\alpha_{em}^2\alpha_s y/(8\pi z_h^2 Q^2)$, $\hat{x}=x_B/x$, $\hat{z}=z_h/z$, and $e_c$ and $m_c$ are the fractional charge and mass of the charm quark, respectively. The spin-averaged cross section $d\sigma$, and the functions ${\cal A}_i$, $\hat{W}_i$ and $\hat{N}_i$ are given 
in Ref. \cite{Kang:2008qh}.
From Eq.~(\ref{dmeson}), we find that the SSA for $D$ meson production is
proportional to $T_G^{(f)}+T_G^{(d)}$. We also calculate the SSA for 
$\bar{D}$ meson production, and find that the SSA for $\bar{D}$
has the same functional form as that for the $D$ meson produciton 
except the sum of the trigluon correlation functions, $T_G^{(f)}+T_G^{(d)}$, 
is replaced by $T_G^{(f)}-T_G^{(d)}$.  That is, the $D$ and $\bar{D}$ 
meson production should have the same SSA if $T_G^{(d)}=0$, but, with an opposite sign if $T_G^{(f)}=0$.  If both trigluon correlations vanish, there
should be no significant SSA for open charm production in SIDIS.

The fully differential cross sections can be decomposed in terms of the independent angular distributions as follows,
\ben
\frac{d\sigma}{dx_B dy dz_h dP_{h\perp}^2 d\phi}&=&\sigma_0^U+\sigma_1^U\cos{\phi}+\sigma_2^U \cos{2\phi}, \nnu
\frac{d\Delta\sigma}{dx_B dy dz_h dP_{h\perp}^2 d\phi}&=&\sin{\phi_s}\left(\Delta\sigma_0+\Delta\sigma_1\cos{\phi}+\Delta\sigma_2 \cos{2\phi}\right).
\een
In order to present the SSA and its angular dependence on the $\phi$,
we introduce the $\phi$-integrated single spin azimuthal asymmetries:
$\langle 1 \rangle = {\Delta\sigma_0}/{\sigma_0^U}$, 
$\langle \cos\phi \rangle = {\Delta\sigma_1}/{2\sigma_0^U}$, 
$\langle \cos2\phi \rangle = {\Delta\sigma_2}/{2\sigma_0^U}$
\cite{Kang:2008qh}.
To estimate these SSAs, we adopt the model for trigluon correlation functions $T_G^{(f,d)}$ introduced in Ref.~\cite{Kang:2008qh}. In Fig.~\ref{zh_dep} we plot the SSAs (only the contribution from $T_G^{(f)}$) as a function of $z_h$ for the COMPASS (left) and eRHIC (right) kinematics. 
The SSA, $\langle 1 \rangle$, is of the order of $10\%$ ($5\%$), 
which could be measurable at the COMPASS (eRHIC) experiment. If the contribution 
from $T_G^{(d)}$ is also included, the SSAs for $D$ meson will be doubled 
if $T_G^{(d)}=T_G^{(f)}$ and vanish if $T_G^{(d)}=-T_G^{(f)}$. 
However, the situation will be opposite for $\bar{D}$ meson. So the difference
between the SSAs of $D$  and $\bar{D}$ production can uniquely separate
contributions from these two trigluon correlation functions, unless both vanish.
\bef
\psfig{file=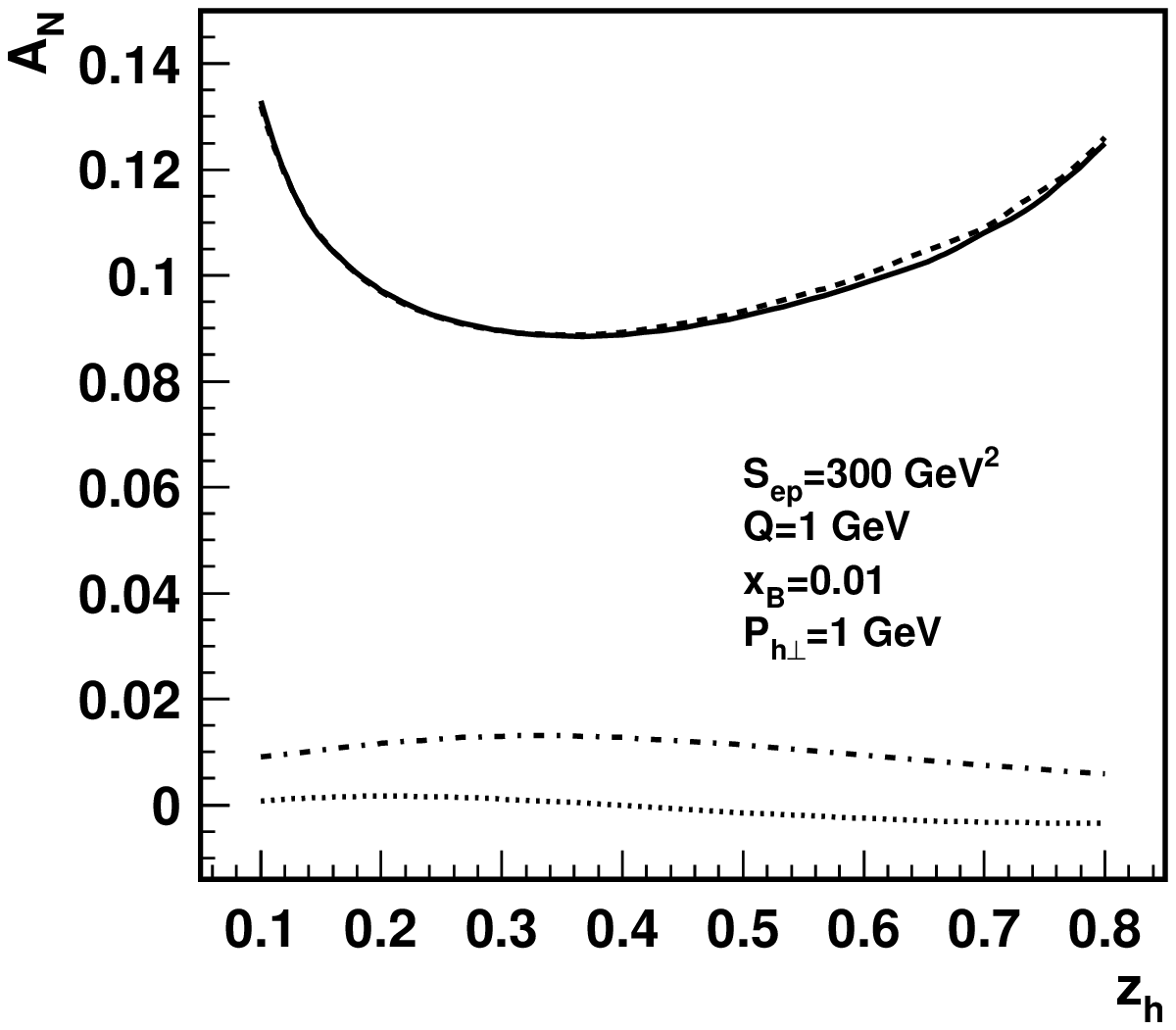,width=2.3in}
\hskip 0.2in
\psfig{file=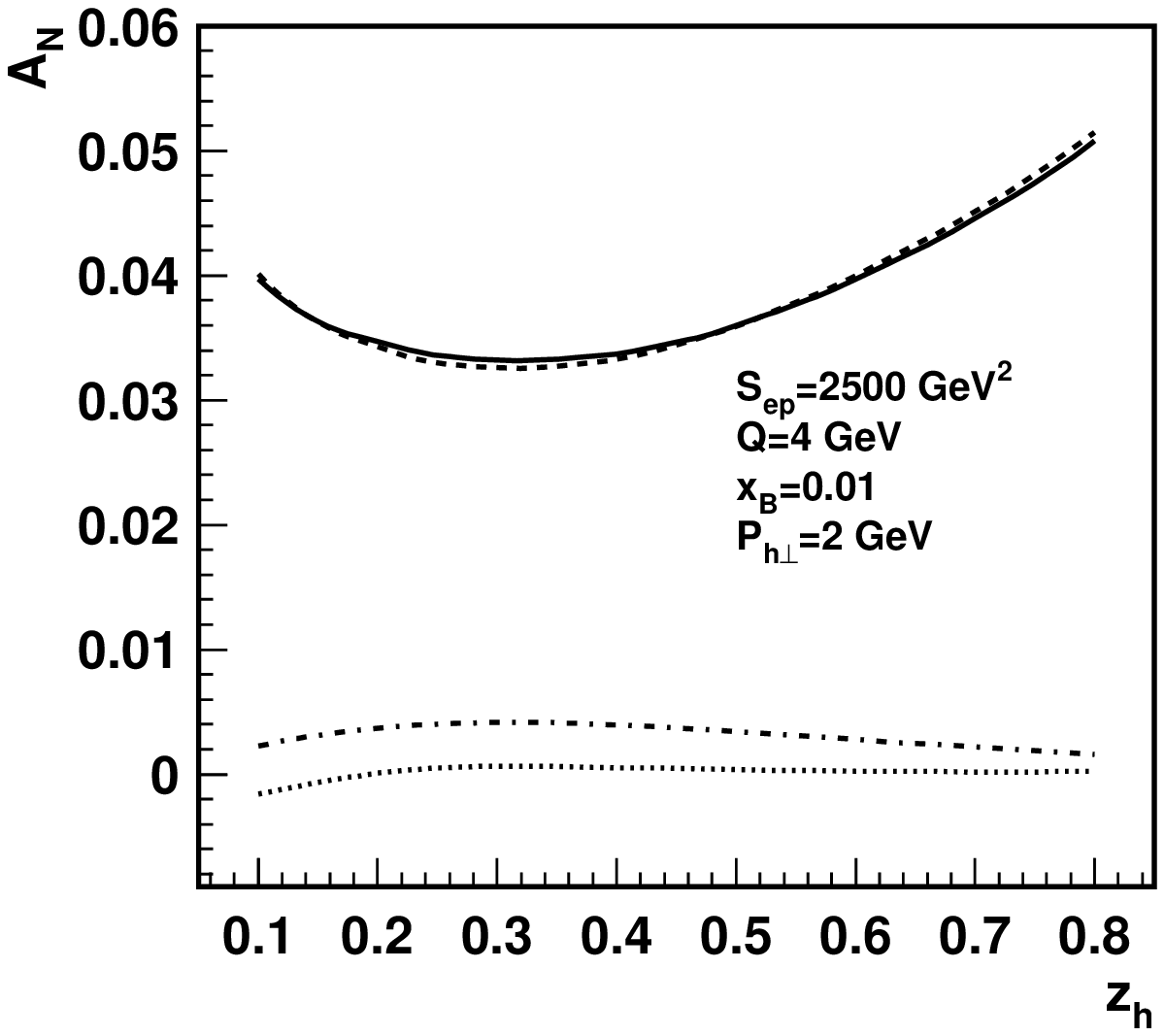,width=2.3in}
\caption{The SSAs for $D^0$ production in SIDIS for COMPASS (left) and eRHIC (right) kinematics. The curves are: solid-$\langle 1 \rangle$, dashed-$\langle 1 \rangle$ with derivative-term only, dot-dashed-$\langle \cos\phi \rangle$, and dotted-$\langle \cos{2\phi} \rangle$.}
\label{zh_dep}
\eef

The SSAs for open charm production in hadronic $pp$ collisions can also 
be calculated similarly. Different from SIDIS process, both quark-gluon correlation function
$T_{q,F}(x,x)$ and trigluon correlation functions $T_G^{(f,d)}(x,x)$
will contribute to the SSAs at the leading order. 
Following the same feature as in SIDIS, the SSAs for $D$ meson ($\bar{D}$ meson) 
will depend on the sum (difference) of two trigluon correlation functions. 
In Fig.~\ref{pt_dep_mid}, we show the $P_{h\perp}$-dependence of the SSAs for both
$D$  and $\bar{D}$ mesons at RHIC kinematics. The different curves correspond to different assumptions
on the two trigluon correlation functions \cite{Kang:2008ih}. 
One can clearly see that the quark-gluon correlation functions 
alone (dashed lines) generate only a very small asymmetry. Therefore, 
the observation of any sizable SSAs would be a clear evidence of 
nonvanishing trigluon correlations inside a polarized proton \cite{Kang:2008ih}.
Due to the different dependence on
the trigluon correlation functions, the asymmetries and the difference 
in the asymmetries of $D$ and $\bar{D}$ meson production could
be powerful observables for extracting both trigluon correlation functions
inside a transversely polarized proton.
\bef
\psfig{file=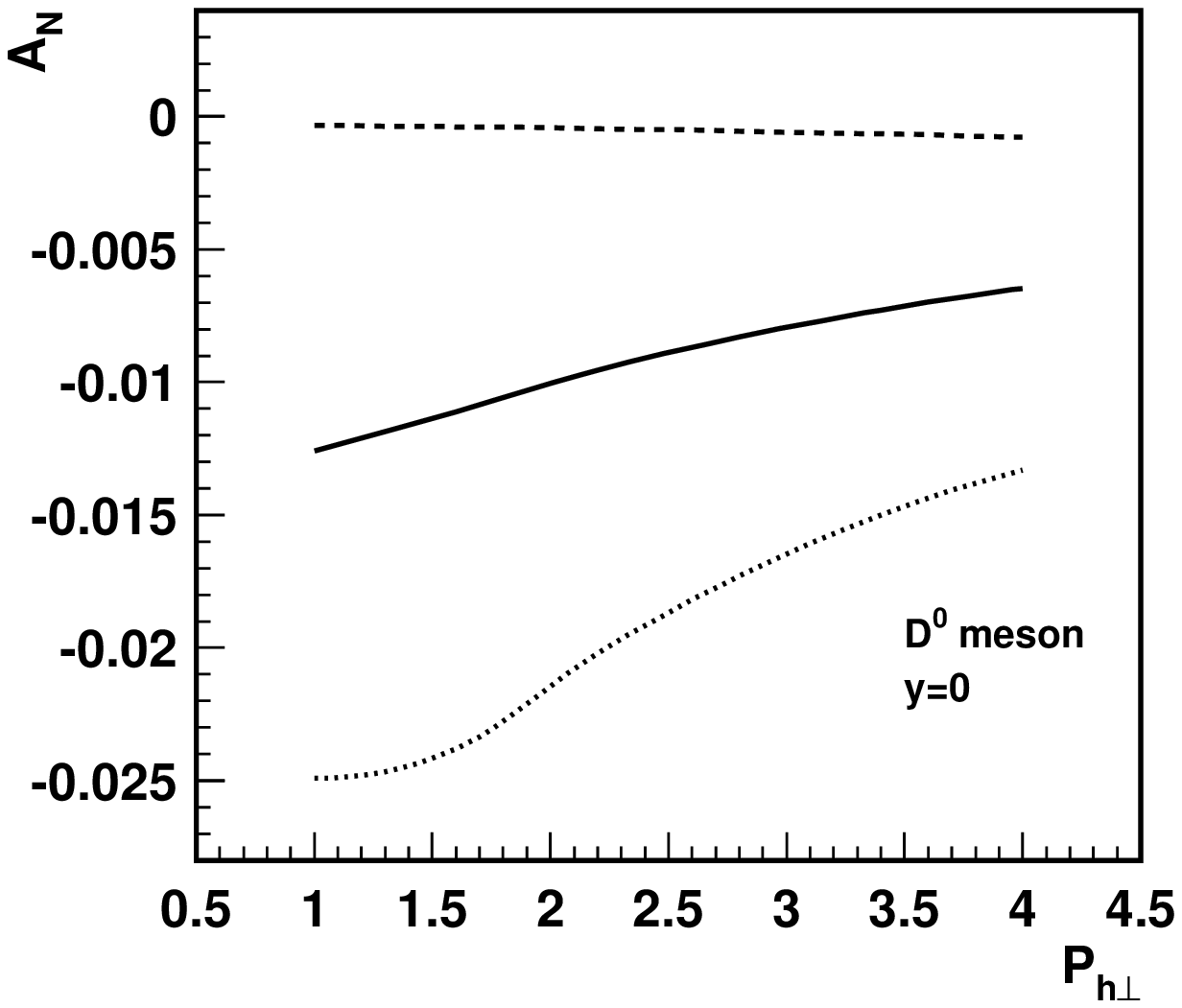,width=2.5in}
\hskip 0.2in
\psfig{file=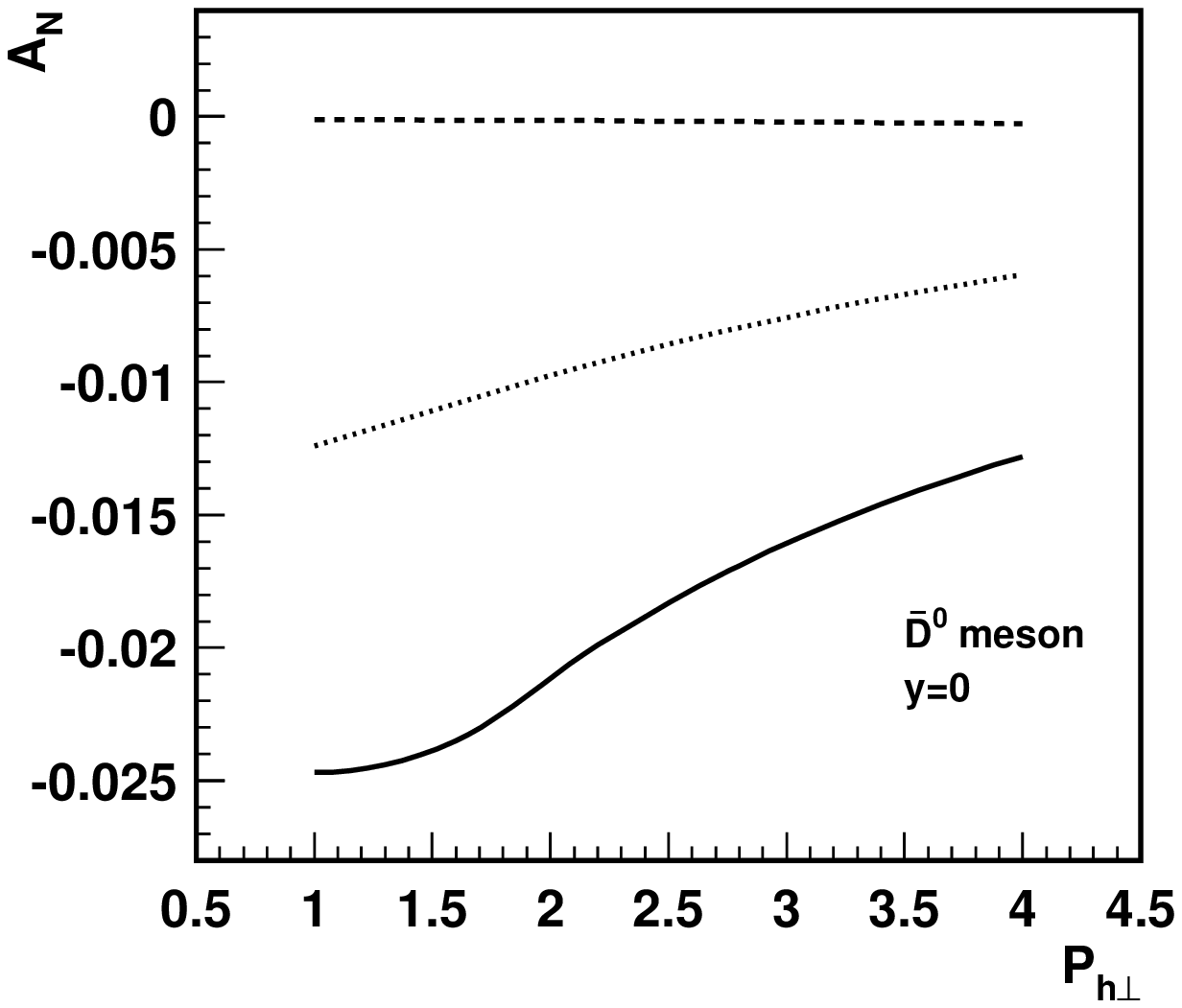,width=2.5in}
\caption{The SSA as a function of $P_{h\perp}$ for $D^0$ (left) and $\bar{D}^0$ mesons (right) at mid-rapidity, $y=0$,
and $\sqrt{s}=200$ GeV. The curves are: solid ($T_G^{(d)}=T_G^{(f)}$), dashed ($T_G^{(f,d)}=0$), dotted ($T_G^{(d)}=-T_G^{(f)}$).}
\label{pt_dep_mid}
\eef

In summary, we have studied the single transverse-spin asymmetries of $D$  and $\bar{D}$ meson production in both SIDIS and $pp$ collisions. We found that two trigluon correlation functions play important but very different role in generating the SSAs for $D$  and $\bar{D}$ mesons. With data on SSAs for open charm production becoming available from 
RHIC~\cite{liu} and perhaps in the future in SIDIS, we will be able to extract 
the trigluon correlation functions and to learn for the first time about
the dynamics of quantum correlations of gluons inside a polarized proton.

{\bf Acknowledgements} We thank Werner Vogelsang and Feng Yuan for discussions and collaborations. We also thank Ming Liu and Han Liu for correspondence on experimental measurements of SSAs in open charm production at RHIC.
This work was supported in part
by the U. S. Department of Energy under Grant No.~DE-FG02-87ER40371.

\end{document}